\def\year{2019}\relax
\pdfoutput=1
\documentclass[letterpaper]{article} 
\usepackage{aaai19}  
\usepackage{times}  
\usepackage{helvet}  
\usepackage{courier}  
\usepackage{url}  
\usepackage{graphicx}  
\usepackage{amsmath}
\usepackage{alltt}
\usepackage{nameref}
\begin{document}
%
\title{Deriving Cyber-security Requirements for Cyber Physical Systems}
\author{Robert Laddaga\textsuperscript{1}, Paul Robertson\textsuperscript{2}, Howard Shrobe\textsuperscript{3},\\ 
{\bf  \large Dan Cerys\textsuperscript{2}, Prakash Manghwani\textsuperscript{2}, Patrik Meijer\textsuperscript{1}}\\
\textsuperscript{1}Vanderbilt University and \textsuperscript{2}DOLL Labs, Inc. and \textsuperscript{3}MIT\\
Nashville, TN and Lexington, MA and Cambridge,MA\\
}
\nocopyright

\maketitle
\begin{abstract}
Today's cyber physical systems (CPS) are not well protected against cyber attacks. Protected CPS often have holes in their defense, due to the manual nature of today's cyber security design process.  It is necessary to automate or semi-automate the design and implementation of CPS to meet stringent cyber security requirements (CSR), without sacrificing functional performance, timing and cost constraints. Step one is deriving, for each CPS, the CSR that flow from the particular functional design for that CPS. That is the task assumed by our system, Deriving Cyber-security Requirements Yielding Protected Physical Systems - DCRYPPS.  DCRYPPS applies Artificial Intelligence (AI) technologies, including planning and model based diagnosis to an important area of cyber security.
\end{abstract}

\section{Introduction}

Embedded and cyber physical systems (CPS) are tremendously important to our military, our nation and the world.  Most of our weapons and military support systems are CPS, much of our critical infrastructure, such as the electric grid, pipelines, and transportation systems are implemented as CPS.  The shift to digital control happened at the same time as our nascent digital computing efforts, and happened like the general computing developments, with little regard for cyber security concerns.  Our critical infrastructure and weapons have always been vulnerable to physical attack, but for such massive systems, physical distribution and the energy required to destroy or incapacitate them was itself a kind of protection.  However, cyber attacks can be inexpensive, and networks can deliver those attacks throughout the world very quickly.  Cyber attacks taking control of CPS control systems have the ability to turn the CPS's native power against itself.

Current systems are vulnerable to attacks, which can lead to catastrophic consequences in a military as well as civil environment.  With the interconnectedness of CPS, such as automobiles, weapon systems and critical infrastructure, this is already a huge problem. With the increasing penetration of the Internet of Things, this will soon become an even bigger challenge.   A core deficiency in the current state of the art is the lack of the internal ability to accurately detect the attack and diagnose the corrupted state.  Current computer systems rely on external host or network intrusion detection systems (IDS) to identify the incidence of intrusions. This approach has a number of intrinsic limitations. Existing IDS can only rely on target system's generic monitoring capabilities, instead of using the application, and system specific information that might be available. Feature  construction (e.g., statistical measures for anomaly detection, attack signatures for misuse detection) has been a challenging issue. Without the knowledge of the systems intended behavior at design time, existing IDS systems rely on learning system run-time behaviors from audit data and extracting their patterns. Without application semantic context, this approach usually results in low detection accuracy and the inability to separate meaningful program state from circumstantial operations.

Furthermore, the lack of application-specific information forces the system administrators to rely on generic mitigation techniques which might end up doing more harm than good. Consider the case where the functionality is  critical and real-time in nature. In this case, any reconfiguration that takes longer than the task's periodicity might result in serious errors. Rather than complete reconfiguration, in this case, a phased migration might be more appropriate  wherein an active replica is started and the state is transferred before the instance under attack is brought down.  Today's CPS are mostly not well protected against cyber-only attacks, let alone hybrid cyber-physical attacks.  Those systems that are internally protected often have holes in their defense, due to the manual nature of today's cyber security design process.

We seek to provide a research driven answer to the problem of how to automate or at least semi-automate the design and implementation of CPS to meet stringent cyber security requirements, without sacrificing functional performance, and timing and cost constraints.  The first step in that process is deriving, for each CPS, the cyber security requirements that flow from the particular functional design for that CPS.  That is the task that our system, Deriving Cyber-security Requirements Yielding Protected Physical Systems - DCRYPPS, takes on.  

\begin{figure}[t]
\centering
\includegraphics[width = 0.7\columnwidth]{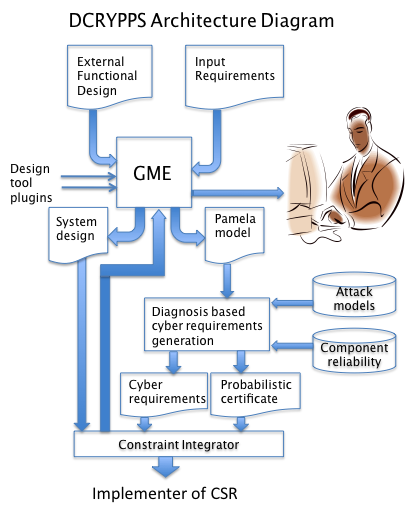}
	\caption{DCRYPPS Architecture Diagram}
	\label{fig:architecture}
\end{figure}

\begin{figure}[t]
  \centering
\includegraphics[width = 0.7\columnwidth]{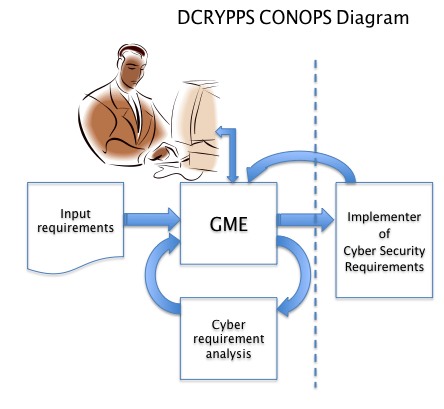}
	\caption{DCRYPPS CONOPS Diagram}
	\label{fig:conops}
\end{figure}

Figure \ref{fig:architecture} shows the architecture of the DCRYPPS system.  The center of the system is the Vanderbilt Generic Modeling Environment (GME) (see \cite{gme}, \cite{ledeczi01thegeneric}, \cite{maroti2014}), which serves as a design environment, as well as a design tool integrator.  Inputs to GME include functional design models and requirements, including constraints on information flow, for example.  One of the design and analysis tools that we have integrated with GME is the DOLL Pamela modeling language, which supports probabilistic models of CPS, plant models, resources and environments, as well as functional descriptions of CPS behavior.  A GME user creates a design, using the input models and requirements, and GME converts that model into a Pamela model, which feeds an analysis tool that uses attack/threat models and component reliability information to drive a diagnosis based analysis of how to thwart the threats.  The outputs of the analysis are cyber security requirements and probabilistic certificates, which are then fed to a constraint integrator.  The constraint integrator provides options, via GME, to allow the designer to modify the tool behavior, for example, by adjusting the tradeoff between risk reduction and the complexity of the output requirements.  Finally, the cyber security requirements are delivered to system designers and implementors responsible for producing a CPS that meets cyber requirements, and well as functional performance, timing and other constraints.  This basic concept of operations for DCRYPPS is shown in Figure \ref{fig:conops}.

\section{DCRYPPS Innovations}

DCRYPPS delivers significant and beneficial innovations, including:

\begin{enumerate} 
\item \textbf{Use of model based diagnosis to derive cyber security requirements from functional design models.} Model based diagnosis is a proven technology for diagnosing a fault given an appropriate model of the system exhibiting the fault.  However, we combine model based diagnostic technology with top down attack models in order to exhaustively map fault achieving threats.  The resulting diagnoses are then used to generate cyber security requirements that, if satisfied, rule out that specific diagnosis. 
\item \textbf{Focus on cyber physical properties and cyber physical attack models.} Many attempts to formalize and prove cyber security properties of systems have focused on information flow, which incorporates confidentiality and modest portion of integrity.  But unlike pure information systems, CPS are physically tied to the world, have potentially dramatic physical effects, and vulnerable not only to cyber attacks, but also physical and hybrid attacks.  We focus on attacks like sensor and actuator signal spoofing, timing attacks and network attacks, where the most serious vulnerabilities of CPS are located. 
\item \textbf{Analysis of cyber requirements that is sensitive to risk and cost tradeoffs.} Although we can generate an exhaustive list of faults, that doesn't mean that we should generate an exhaustive list of cyber requirements.  Our system orders consideration of threats by likelihood and significance of the effect of the threat.  We can then tradeoff how deeply we descend the list of threats, against the remaining risk.  We give system designers tools to explore this design tradeoff space, and the ability to adjust cost and likelihood metrics.
\item \textbf{Use of probabilistic certificates in managing cyber security requirements.} Probabilistic Certificate of Correctness (PCC) is a metric used to capture risk in the engineered
systems development processes.  We use the PCC as a way of
validating that the cyber requirements cover the needs to avoid cyber
attacks. To achieve the maximum benefit from
this approach it is important that a requirements verification process
with a PCC metric is implemented in a consistent way for all
requirements. 
\end{enumerate}

\section{DCRYPPS Technical Details}


Building on Vanderbilt's history in Cyber Physical Systems (CPS), we leverage our tool GME / Meta-GME for initially specifying the system.  GME has had decades of use in both designing CPS, and in accepting a variety of modeling languages and methodologies as inputs.  One of our first tasks was to integrate DOLL's Pamela probabilistic modeling language with GME, so models built or imported into GME can be exported to Pamela, where they can be analyzed using Pamela's suite of solvers.

We treat the derivation of cyber security requirements as a kind of inverse diagnosis problem.   We use attack and threat models as guide for injecting faults into functional models, and use diagnostic reasoning to derive cyber security requirements that can effectively block those specific diagnoses.  This combination of functional system models and attack models was first explored in Shrobe's MIT AWDRAT Self Regenerative Systems project \cite{AWDRAT}, and builds on prior model based diagnosis work \cite{shrobe-iwsas2001}, \cite{shrobe-2002b}, and \cite{davis-shrobe-82}.  We carry this approach further by using it not just to elaborate attack models, but to actually derive cyber security requirements.

The inputs to the system are a functional model of CPS, in a language such as AADL (see \cite{feiler_AADL:06}, \cite{AADL_Intro:06}), SYSML (see \cite{sysML_v1.3:12}), or other model based design language, along with a set of desirable properties that designers believe must be maintained (invariants) even in the face of cyber attacks.  An additional input is the level of confidence the designer want to have that the invariants will hold. 

Essentially, in this approach, we hypothesize various invariant violations that could occur, by negating each of the desirable property invariants.  We assert that the violation has happened and we run diagnosis to generate possible causes.  Then for each cause, we ask of the component implicated if various attack models are applicable to that component, and if so, we generate a cyber requirement such that that component be resilient against that attack model.  Having added that requirement, the component is deemed safe and hence the problem must be elsewhere.  We iterate through the possible causes in order of most likely first until we reach the point where there are no more possibilities that are cyber; hence in the end, the diagnosis will always be the failure of a part due to non-cyber causes.

Given a risk target, we avoid descending too far down the list of decreasingly low probability chain of possible requirements so as to manage the overall size of the cyber requirements list. We can also weight the acceptable risk based on the importance of the function so that more cyber requirements are generated to protect the most important assets.  We keep track of the accumulated risk so that the generated cyber requirements should be verifiable by an implementer to produce a certificate within the acceptable risk specified.

Finally, we emit the accumulated cyber requirements which essentially are constraints on the original system design.

\subsection{Use Case}
\label{ssec:usecase}

For a worked use case, we consider a simple autopilot for a remotely piloted quadcopter.

The quadcopter is guided from a ground station using cellular
networking. The ground station can send new waypoints to the
quadcopter but the quadcopter always has a default plan to follow in
case the communication with the ground station is lost.

\begin{figure}[t]
\centering
\includegraphics[width = 0.8\columnwidth]{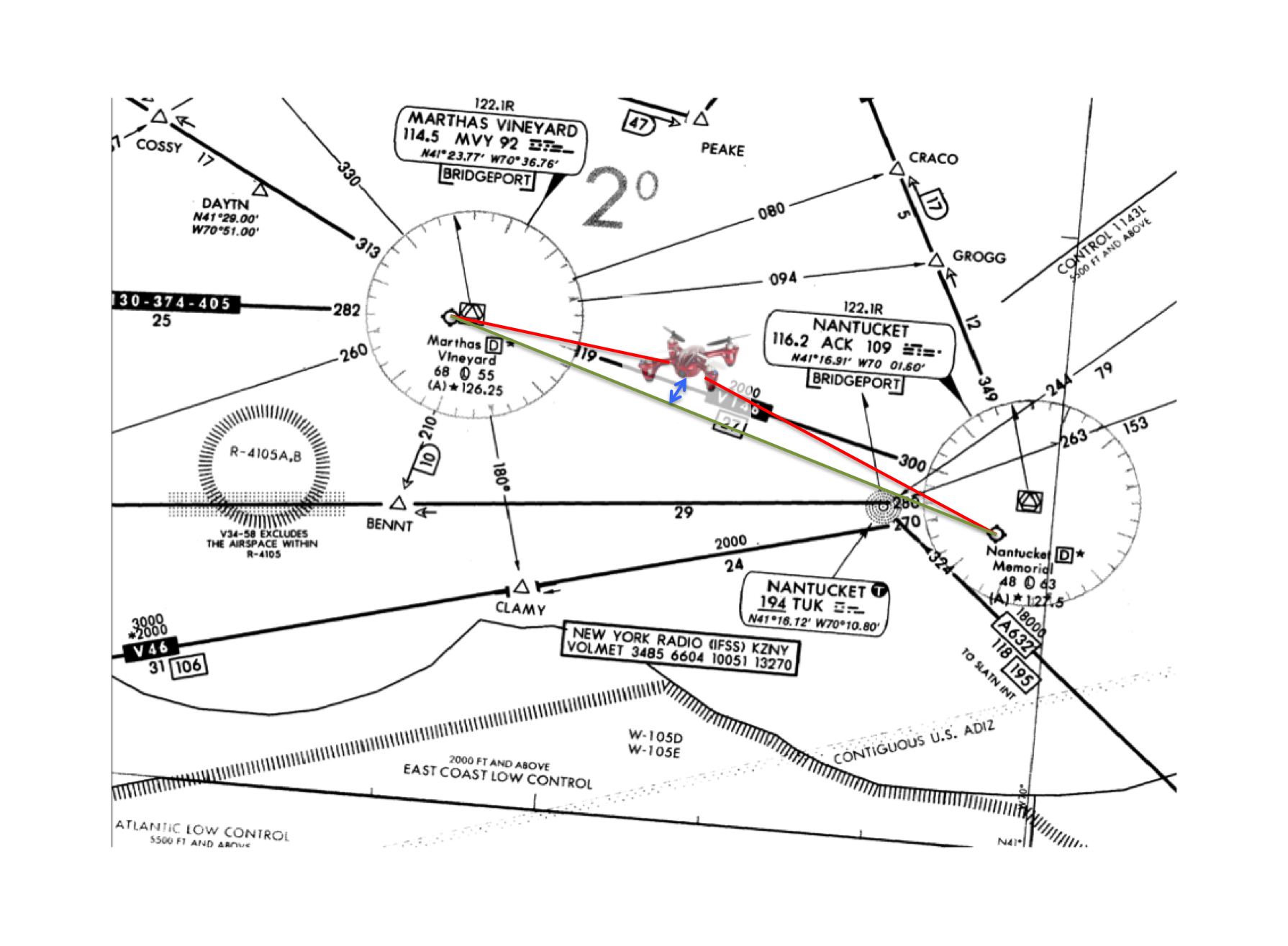}
	\caption{Auto pilot scenario.}
	\label{fig:autopilotscenario}
\end{figure}

It does, however, require reliable navigation.  For this reason, the
autopilot, whose job it is to follow waypoints until landing at the
final waypoint, has two different sensors, a (Global Positioning
System) GPS receiver which gives position and altitude, and a VHF
Omnidirectional Range (VOR) sensor that gives directions towards a
ground based transmitter.  With the aid of navigation maps that show
the locations of VOR transmitters, it is possible to navigate between
VOR transmitters. Figure~\ref{fig:autopilotscenario} depicts a
quadcopter en route from Nantucket to Martha's Vineyard.

The job of the autopilot is to follow the path between waypoints.  It
uses sensors to determine position, it uses a Kalman filter to estimate
position, calculates an error term and then calculates the appropriate
control signals to feed to the actuators (which are themselves simple
controllers).  This is susceptible to all the types of attacks.

The command interface has a web server for use at the ground station.
Everything is controlled by the autopilot controller board. All of of
the sensors, the GPS and the VOR, communicate with the controller
board on a local network. The VOR requires an interface that sets
frequencies and directions and the GPS has an interface that provides
information on satellites being received and position. The autopilot
has an up to date listing of all VORs in its operating region but the
GPS is the preferred sensor for a variety of reasons that we won't go
in to here.

\begin{figure}[t]
\centering
\includegraphics[width = 0.8\columnwidth]{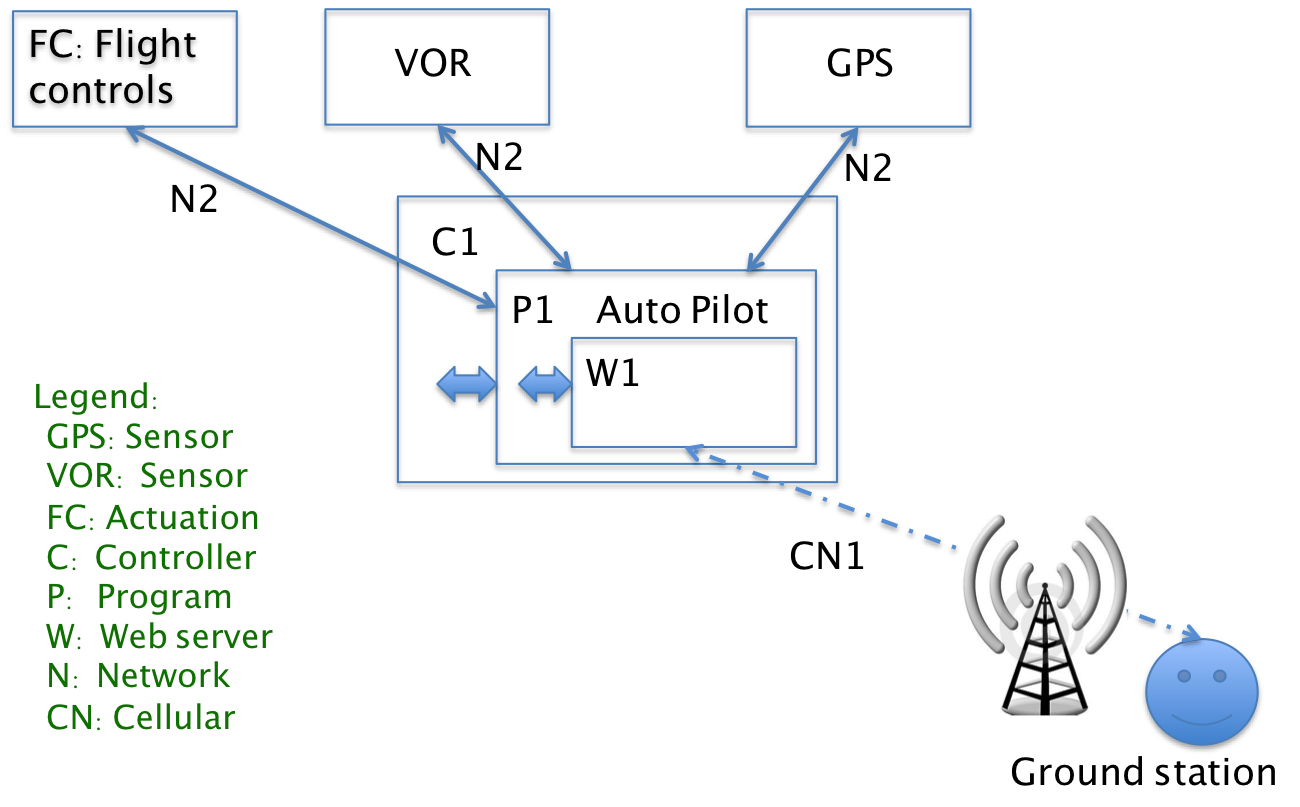}
	\caption{A simple autopilot.}
	\label{fig:autopilotsystem}
\end{figure}

The model of the autopilot is shown in Figure
\ref{fig:autopilotsystem}. For simplicity we have elided many details
including the Kalman filter algorithm used by the autopilot program to
track position estimate independently for the two sensors VOR and GPS.
The controller board, C1 supports the autopilot program, P1, which
communicates with a local webserver W1 which in turn communicates with
the ground station via the cellular network, CN1, in order to
communicate with the ground station client.  The two sensors VOR and
GPS communicate with P1 over a local network. The autopilot program
continuously maintains an estimate of 'position' using its two
sensors. It maintains also 'distance-from-trajectory' which is an
estimate of how close to the target trajectory the quadcopter is at
each moment. Other variables such as 'distance-to-next-waypoint' are
also maintained.

The designer also defines a set of propositions that define what
should not occur in the system, the invariant violations. Some of
these propositions can be proposed automatically given the
components being used, others are manually added to by the designer to
indicate what is essential.

\textbf{Violation of invariants:}
\begin{equation}
\begin{split}
distance(VOR,GPS)>\\     MaximumSensorDisagreement \\
distance-from-trajectory>\\   MaximumOffTrajectory \\
...
\end{split}
\label{eq:autopilotbadthings}
\end{equation}

Given this model, developed in GME or another system design tool for
which a plug in is available for GME, the nature of all of the
resources resides in resource models.  In this example we have
standard parts, the GPS and VOR sensors in the
$Hardware>Sensors>Navigation$ resource hierarchy, and the control board
in the $Hardware>Computer>Controller$ resource hierarchy; the autopilot
program P1 and the web server are in $Software>$ ... and so on.  Each
resource model contains details of all aspects of the component in
terms of requirements for its proper function, such as power and
memory and also its cyber characteristics.

The structure of the system design, which is essentially an attributed
UML diagram, is translated by GME into Pamela to enable the cyber
requirement diagnosis. a rough sketch of which is shown below.

\begin{alltt}
;; Many details including inheritance 
;; of resource attributes omitted
;; in this example for clarity.
(defpclass VOR [localnet]
  ...)
(defpclass GPS [localnet]
  ...)
(defpclass FlightControls [localnet]
  ...)
(defpclass ControllerBoard 
                 [localnet cellnet]
  :fields (pclass AutoPilotProgram 
                 self localnet cellnet)
  ...)
(defpclass AutoPilotProgram [controller 
                    localnet cellnet]
  :fields (pclass webhost controller 
                   localhost cellnet)
  ...)
(defpclass WebServer 
           [board localnet cellnet]
  ...)
(defpclass Network [...]
  ...)
(defpclass CellularNetwork [...]
  ...)
;;; This class wires components
(defpclass AutoPilotUnit []
  :fields 
  { :n2 (lvar ``localnetwork'')
    :cn1 (lvar ``internet'')
    :controlboard (lvar ``cb'')
    :gps (pclass GPS  :n2)
    :vor (pclass VOR  :n2)
    :fc (pclass FlightControls :n2)
    :localnet (pclass Network 
          :controlboard :n2 ...)
    :cellnet (pclass CellularNetwork 
           :controlboard ...)
    :controller (pclass :controlboard 
           :localnet :cellnet) }
  ...)
\end{alltt}

Given the Pamela model, the cyber requirements generator invokes the
diagnosis engine multiple times each time asserting one or more violation in order to produce a list of vulnerabilities.  In this very
simplified example, diagnosis will be invoked in order on:

\begin{equation}
\begin{split}
dist(VOR,GPS)>MaxSensorDisagreement
\end{split}
\end{equation}

\begin{equation}
\begin{split}
distFromTraj>MaxOffTraj
\end{split}
\end{equation}

\begin{equation}
\begin{split}
(dist(VOR,GPS)>MaxSensorDisagreement \land\\ distFromTraj>MaxOffTraj)
\end{split}
\end{equation}

The diagnosis, which is described in the Section titled \nameref{diagnosissection},
uses the structural nature of the system model to identify parts that
may be vulnerabilities that cause the violations to occur. These
things include normal failure as well as cyber attack possibilities.
Normal failure probability is determined by parameters of the resource
model for the parts, and cyber attack possibilities are generated from
matching cyber attack models to the violations and the components
involved in the system design that are connected to the violations
in question.

In this case, $distFromTraj>MaxOffTraj$
implicates in various degrees, the position sensors, the network over
which they communicate, and the autopilot program.

One of the vulnerabilities detected by the diagnosis engine
is a match with 'spoofing attack'. Different variations of the spoofing
attack are revealed: spoofing of the VOR sensor and spoofing of the
GPS sensor; and man-in-the-middle attack on the flight
controls is also identified.  Each of these cause cyber requirements
to be generated for the components involved in the identified attack
vulnerabilities, namely the auto pilot program itself, the local
network, and the cellular network.

\subsection{Attack Modeling}
\label{sec:attackmodeling}

We use cyber attack and threat modeling in order to assist in driving the diagnostic analysis that examines components and connections to track and eliminate faults.
In particular, an essential component of the proposed system is an attack matcher which will be invoked by the vulnerability finder as part of the cyber requirements generation.  A database of attack types will support potential matches against part of the system design that is under review.  It responds to the question: given that an invariant violation has been asserted, is there an attack type in the database that could result in that violation?  A match will result in a possible cyber requirement to present that attack from succeeding.

We begin by considering several scenarios that arise in embedded  control systems.  In their simplest form, these control systems have four main elements:
\begin{enumerate} 
\item A set of sensors, with analog signals that are converted to digital form, collected in a data concentrator and then transmitted to the control element
\item A network, that in modern systems transmits both sensor data and control signals, using Ethernet cabling and lower level protocols, but preserves the older MODBUS or FIELDBUS application level protocols
\item The controller per se, that receives sensor data,  estimates the state of the system, compares the estimated state to the desired state, and based on the difference between the two, computes control signals that are transmitted to actuators
\item Actuators (or effectors) are digitally controlled devices that, in response to the control signal, have physical effects on the system under control.
\end{enumerate}

For the system to work correctly: sensor data must be captured and transmitted to the controller without modification; 
the controller's code must continue to accurately estimate the system state and correctly compute the transfer function (state error to control signal);
the control signal must reach the actuators unchanged;
the actuators must respond  correctly to the control signals;
and each cycle from sensor to controller to actuator must take place within a time interval that depends on the physics of the system under control.

Beneath this level of abstraction, each of these actions are carried out by computers: 1) Sensor data is captured by a data concentrator, a specialized but usually programmable computer. 2) The controller, often referred to as a Programmable Logic Controller (or PLC) is typically a programmable computer built around a micro-controller processor (or even a standard MIPS or ARM chip). 3) All elements interact with the network through a NIC (network interface card), which is usually built from a SoC (system on a chip).  The firmware in these elements historically was in ROM (i.e. unwritable), but in more modern systems the firmware is kept in EEproms or other writable elements.

Finally, note that the controller unit is today implemented as a standard computer, running a standard Real-time OS with the PLC firmware running as a real-time process.  In addition to the PLC firmware, most modern systems provide a web server (usually Nginx or Apache) that provide the management interface to the system.  Because, it is used as the management system, the web server runs at a privilege level that allows it to modify critical elements of the controller system (including setting of process control parameters and installing new control programs to be executed by the controller).

Given this and attacker goals to, for example, disrupt the system, we can derive a number of attack scenarios, below.

\subsubsection{Sensor Data Spoofing}

If an attacker can modify all sensor data that reaches the controller, then a perfectly correct controller will issue actuator commands that can have deleterious effects.  For example, imagine a temperature sensor whose values are changed to read low.  The controller will respond by commanding the actuators to provide more heat (say by increasing fuel or oxygen supply).  Sensor spoofing is a very pernicious form of attack since the controller's only perception of the real-world is via its sensors.

Sensor spoofing can be effected in either of two general approaches: a) By penetrating the data concentrator and then using it to systematically change the real sensor data to spoofed data.
b) By man-in-the-middle attacks on the network linking the data concentrators to the controller.

\subsubsection{Modification of the Control System's Transfer Function}

There are several ways an attacker can do this:

a) In systems where the management interface is provided by a web server (or some other remote user interface) any of the vulnerabilities of the web server might be exploited to change control parameters or to download hacked versions of the controller code (as was done in Stuxnet, although not via a web interface).

b) There might be sensitivities in the controller code to certain numeric values (or sequence of values) that can induce numeric errors that lead to unintended control flows, code injection, or code reuse attacks.  

c) The controller maintains real-time network connections to its data concentrators and to its supervisory and human interface systems.  If any of these network protocols can lead to buffer overflows, then these can be exploited to inject  code or to conduct code-reuse attacks.

d) The controller runs an off the shelf real-time OS.  If this leaves certain ports open (e.g. FTP or Telnet) then the attacker may be able to log in via a stolen or guessed password and then launch programs that can affect the controller process.

\subsubsection{Timing Attacks}

The sense-compute-actuate loop must execute within the time constant of the system under control.  Anything that the attacker can do to change the timing can have disastrous effects.

1) Depending on the scheduler of the system, excessive load placed on processes other than the controller can cause the controller process to miss its deadline.  

2) If it's possible to initiate a remote login, then this can be used to launch a large number of jobs.  In the worst case, these would be jobs with real-time requirements more demanding than that of the actual controller

3) The network can be saturated enough to make the sensor to actuator loop take too long.  This could be done with a standard DDoS type of attack, although more subtle attacks might well be possible.

\subsection{Generating Cyber Requirements Through Diagnostic Analysis}
\label{diagnosissection}

Given a model of the system that described all of the connections,
dataflows, compute elements, and other attributes of the system design,
we can use diagnosis~\cite{deKleer1987} to find the cause of a fault.  This
technology is useful for diagnosing faults at runtime, but it can be
used also to find possible causes for hypothetical failures of the
system at design time.

First of all, our model needs to represent failure of the system, for whatever reason.  A CPS can fail
due to a hardware failure such as a failed sensor.  It can also fail
because of a cyber attack.  These are often difficult to distinguish,
but our goal is to generate cyber requirements that render the cyber
attacks impossible.  The resulting diagnosis will therefore generate
system failures because the cyber possibilities will have been handled
by the cyber requirements.

A good abstract explanation of how diagnosis works, can be found in
~\cite{deKleer1987}.

Violation implies that something has gone wrong.  Sometimes this is
because of a failed part, sometimes it can be the result of a cyber
attack. We can provide qualitative levels of violation so that the more
serious violations can be given higher priority depending on what is
demanded in the probabilistic certificate.

Example of qualitative levels of failures may be:
\textit{Catastrophic},
\textit{Reduced Capability},
and
\textit{Annoyance}.

The reverse diagnosis works by asserting that each one, in turn, of the
violations has occurred and then diagnosing the fault. After all single
fault cases have been diagnosed, dual fault cases are iterated through
and so on.

For each diagnosis step, the model is inspected to find which parts
could contribute to the failure.  It generates a list of possible
reasons for the failure and rank orders them in order of likelihood.
Starting from the most likely causes, if the cause is a cyber attack,
a cyber requirement is generated and attached to the model component
in question.  That cause from that component will never be listed as
having this vulnerability, so when we meet the same component from a
different diagnosis, it will not be listed as a cyber requirement
candidate.

By enumerating the faults in order of likeliness and by keeping the
causes in order of likelihood, we can add the cyber requirements in
order of greatest impact.  Given a specified acceptable risk, we can
stop adding cyber requirements when the acceptable risk level has been
achieved by the cyber requirements.

The values for likelihood and impact are computed using the attack
models and the system model. 

The computation of the probabilities comes from rules in the attack
models for cyber related failures or in the component models for
failure rate based failures. In this case, for example, a hardware
failure, such as a faulty memory chip in the control board could
result in such a fault. For the non cyber attack faults, the numbers
can be estimated by standard Mean Time Between Failures (MTBF)
testing, and similar results can be obtained from software testing.
These numbers could be augmented by machine learning from in situ data
acquisition given the system model.  In addition, the probabilities of
cause can be adjusted based on distance in steps from the
fault~\cite{LaddagaRobertson1}.

\subsection{Probabilistic Certificates}
\label{sec:probcerts}

Probabilistic Certificate of Correctness (PCC) \cite{PCC} is the manner in which
we will formally encapsulate a model of cyber risk.  The risk model
establishes overall risk on the basis of hierarchical decomposition of
the design and thereby an accumulation of risk fragments throughout
the design.

The system definition is a connection of hierarchically nested parts.
This hierarchical decomposition enables the calculation of a PCC for
system requirements. PCC is a metric used to capture risk in the
engineered systems development processes, in this case we use the PCC
as a way of validating that the cyber requirements cover the needs to
avoid cyber attacks. In a transitioned system, the PCC would include
non cyber requirements as well and the PCC is easily extendable to
this, but for the purposes of this program we will limit ourselves to
cyber requirements. To achieve the maximum benefit from this approach
it is important that a requirements verification process with a PCC
metric is implemented in a consistent way for all requirements.  Our definition of the PCC will include the following key
elements:

\begin{enumerate}

\item Probability of satisfying the cyber requirements (Ps), which
  gives the expected behavior as an estimated probability for a given
  confidence level (as a project risk parameter).

\item Confidence in the probability of satisfying cyber requirements,
  which gives a statistical confidence in the estimated
  probability. Confidence relates to the likelihood of a given
  component of the system being compromised with an attack type given
  the provided safeguard against that remedy.  Each potential
  vulnerability of the system will be analyzed and where appropriate
  constrained by a cyber requirement.
\end{enumerate}

\subsection{Pamela}
\label{sec:pamela}

The Pamela language is a system modeling language that supports
reasoning over systems and machine learning of probabilistic
variables.  Pamela builds upon a long history in model-based
programming.

The modeling of processes has a long tradition rooted in pioneering
work done at Xerox PARC~\cite{xparcref1}, NASA~\cite{remagref1}.  The
work has historically been used to support automated reasoning about
systems and missions, to plan for mission implementations, and to
dispatch the resulting plans, monitor their progress, diagnose and
track model state at run time. The success of these models, developed
over a period of 15 years, has lead to a rich understanding of modeling
languages.  These languages have typically been connected to reactive
planners~\cite{burtonref1}, diagnosis engines, temporal planners
~\cite{kirkref1}, and dispatchers. These models have been compiled into
specialized representations supporting specialized solvers, such as
Temporal Plan Networks~\cite{tpnrefs} (TPN) for the temporal planner,
Simple Temporal Networks~\cite{stnrefs} for the dispatcher, and
Probabilistic Hierarchical Constraint Automata~\cite{phcaref} for
diagnosis.  These specialized representations have in turn been the
focus of specialized algorithms with good performance
characteristics~\cite{algref1,algref2}.

The most recent implementation of one of these languages, Reactive
Model-Based Programming Language~\cite{RMPLrefs} (RMPL) comes at the
tail of a long history of programming language design for capturing
process algebras, most notably RAPS~\cite{RAPSref1},
and Esterel~\cite{esterelref}.

Pamela is a language that supports the modeling process to
produce a generalized process model that can be refined into a
collection/composition of specialized models for which there exist
specialized solvers. As a probabilistic programming
language, Pamela is back end agnostic, and allows the
programmer to specify his program in a language that generalizes over
the known ML landscape so that the back end can reason about the best
ML techniques and solvers for the model and the available hardware.

\subsubsection{Nature of the language}


The modeler frequently wants to define nested complex objects with
complex interaction.  Pamela is therefore an
Object Oriented Domain Specific Language (OODSL).

Pamela provides probabilistic variables, as
  well as state variables, mode variables, finite domain variables,
  and distributions over each.


\begin{figure}[t]
\centering
\includegraphics[width = 0.9\columnwidth]{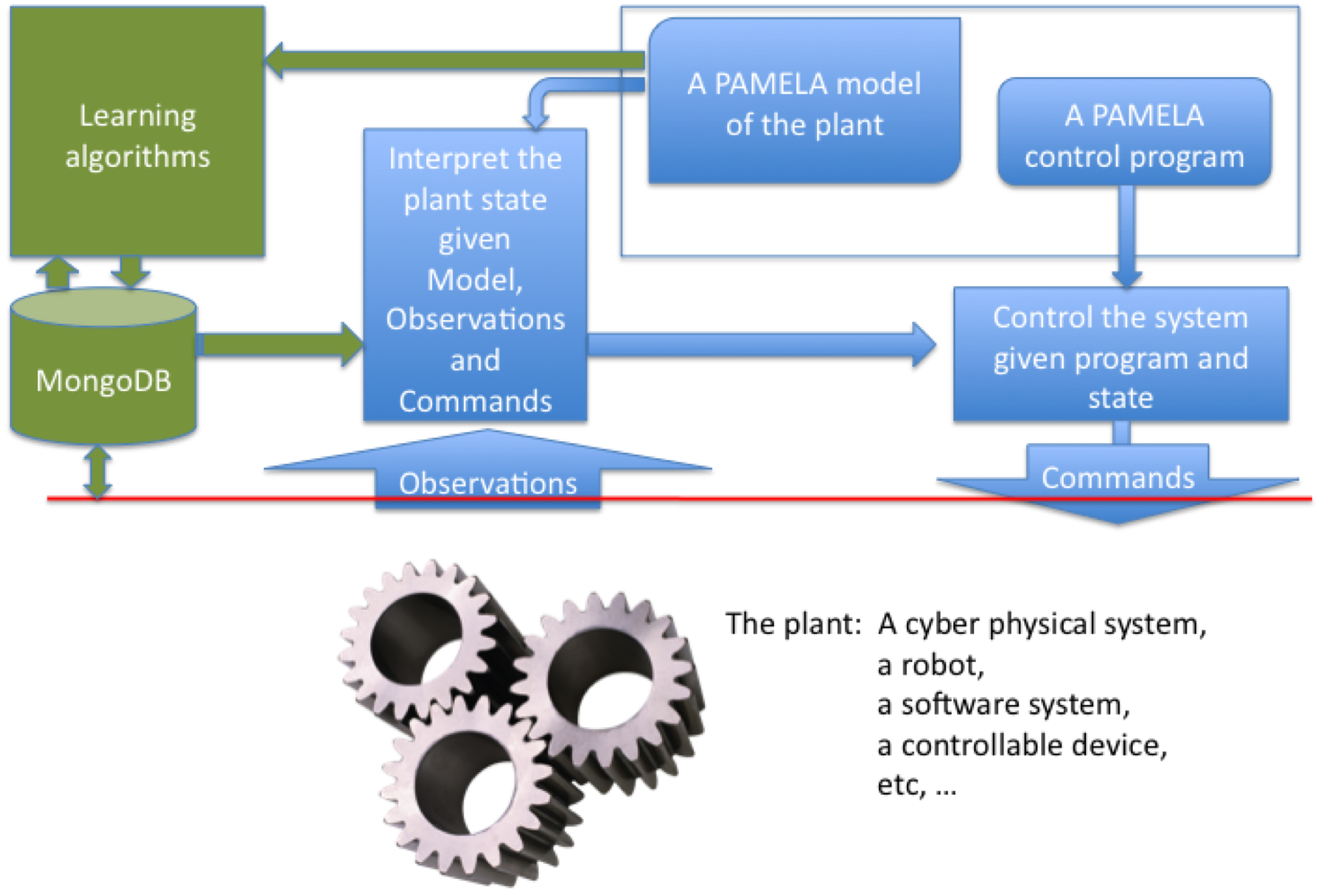}
\caption{Pamela Architecture}
\label{fig:pla}
\end{figure}

Pamela is an actively developed open source modeling language that builds upon RMPL in many ways, but most significantly by adding
probabilistic variables, in situ data collection, and availability as an open source project.

The general architecture of a model based program for a cyber physical system in Pamela is shown in blue in Figure~\ref{fig:pla}.

The plant, a CPS, a robot, a software system, or a controllable device, is
monitored by the model-based program that uses observations and
commands along with the model of the system to predict modeled parts of
its state.  A control program in the context of the plant state causes
commands to be emitted that achieve desired states in the plant.  In
DCRYPPS we use the model offline in order to reason
over cyber requirements of the system design, but it should be noted
that the same model used in the delivered system can also monitor and
track the cyber state of the system and could be used for validation purposes.

The Pamela Learning Architecture adds learning, shown in blue in
Figure~\ref{fig:pla}, and brings in support for data collection and
learning algorithms that can learn the bindings of the probabilistic
variables.  Given that we are generating PCCs as part of this
effort, it is important to ask where the probabilities come from.  For
the non cyber part of the system, measuring the probabilities in real
systems is an interesting possibility and with the Pamela architecture, the
in situ learning of probabilistic variables could be used for
accurate probabilistic estimates.  The situation for cyber attacks is
more complicated given that at any moment a new form of attack could
be invented.  Since we are unable to reason over attack types that
have yet to be seen, we can only reason over known attack models and
measure the effectiveness of the solutions in hosted systems.

\section{Evaluation}

DCRYPPS was evaluated on an autopilot architecture similar to the use case presented here.  The autopilot model, implemented in Pamela,
was accompanied by a set of 17 desirable properties.  DCRYPPS generated five cyber physical requirements.  An independent evaluation team determined that our five cyber requirements were necessary and sufficient to protect the autopilot CPS given the agreed terms of engagement.

The agreed upon terms of engagement were that the attacker does not have physical access to the system at any point in its life cycle; 
the attacker does not have ability to modify the hardware or software during its development or deployment; 
the attacker is assumed to have complete knowledge of the system design, software, and its memory layout;
the attacker has remote access to the system through the internet and radio.

Categories of desirable properties included: safety, system protection, performance, regulations, resources and
information security.

A selection of desirable properties that contributed to the generation of output cyber requirements were three properties specifying that the absolute value of position discrepancies between pairs of position sensors (VOR, GPS and inertial navigation) be below specific threshholds.

An example output cyber requirement was:

WAN (Cellular) communication between Ground Station and Autopilot should be
authenticated using public key encryption.

\section{Conclusion and Further Work}
Given a small scale formal model of a CPS, and a set of desirable invariant properties, DCRYPPS was able to generate an adequate set of cyber requirements.  Future work includes completing the development of PCCs to accompany the generated requirements, and scaling up
to larger CPS.  Also needed is the ability to read in and convert to Pamela models represented in other modeling languages.

\section{Acknowledgment}
This material is based upon work supported by the Defense Advanced Research Projects Agency (DARPA) and Space and Naval Warfare Systems Center Pacific (SSC Pacific) under Contract No. N66001-18-C-4005.

\bibliographystyle{aaai}
\bibliography{bib,f6,pomdp,diag,main}

\end{document}